\documentclass[11pt]{article}

\usepackage[utf8]{inputenc}
\usepackage[T1]{fontenc}
\usepackage{graphicx}
\usepackage{amsmath,amssymb}
\usepackage{booktabs}
\usepackage{hyperref}
\usepackage[margin=1in]{geometry}
\usepackage[numbers,square,sort&compress]{natbib}
\usepackage{xcolor}
\usepackage{caption}
\usepackage{subcaption}
\usepackage{float}

\title{\textbf{Reddit Deplatforming and Toxicity Dynamics on Generalist Voat Communities}}

\author{
  Aleksandar Tomašević\textsuperscript{1} \and
  Ana Vranić \textsuperscript{1} \and
  Aleksandra Alorić \textsuperscript{1} \and
  Marija Mitrović Dankulov \textsuperscript{1}
  \\[0.5em]
  \textsuperscript{1}Institute of Physics Belgrade, University of Belgrade, Pregrevica 118, Belgrade, Serbia
}

\date{December 2025}

\begin{document}

\maketitle

\begin{abstract}
Deplatforming, the permanent banning of entire communities, is a primary tool for content moderation on mainstream platforms. While prior research examines effects on banned communities or source platform health, the impact on alternative platforms that absorb displaced users remains understudied. We analyze four major Reddit ban waves (2015--2020) and their effects on \textit{generalist} communities on Voat, asking how post-ban arrivals reshape community structure and through what mechanisms transformation occurs. Combining network analysis, toxicity detection, and dynamic reputation modeling, we identify two distinct regimes of migration impact: (1) \textit{Hostile Takeover} (2015--2018), where post-ban arrival cohorts formed parallel social structures that bypassed existing community cores through sheer volume, and (2) \textit{Toxic Equilibrium} (2018--2020), where the flattening of existing user hierarchy enabled newcomers to integrate into the now-dominant toxic community. Crucially, community transformation occurred through peripheral dynamics rather than hub capture: fewer than 5\% of newcomers achieved central positions in most months, yet toxicity doubled. Migration structure also shaped outcomes: loosely organized communities dispersed into generalist spaces, while ideologically cohesive groups concentrated in dedicated enclaves. These findings suggest that receiving platforms face a narrow intervention window during the hostile takeover phase, after which toxic norms become self-sustaining.
\end{abstract}

\textbf{Keywords:} deplatforming, content moderation, platform migration, online communities, toxicity, social network analysis, alt-tech platforms

\section{Introduction}

Social media platforms face a fundamental tension between enabling open discourse and preventing harmful content. When communities persistently violate platform policies, platforms may resort to \textit{deplatforming}: the permanent removal of entire communities \citep{chandrasekharan2017cant, jhaver2021evaluating}. Studies confirm that deplatforming reduces hate speech on source platforms \citep{chandrasekharan2017cant}.
However, this framing overlooks a critical dimension: the fate of platforms that absorb displaced users.

In response to stricter moderation, ``alt-tech'' platforms have emerged. These are structural clones of mainstream platforms differentiated by permissive content policies \citep{trujillo2024measuring}. Voat.co exemplified this phenomenon: a Reddit clone launched in April 2014 that positioned itself as a ``free-speech alternative'' with minimal content moderation \citep{mekacher2022voat}. Despite becoming known as a destination for banned Reddit communities, Voat also hosted diverse content including many generalist communities (such as /v/funny, /v/gaming, /v/technology) with relatively benign topics compared to the platform's more extreme spaces. Voat's growth was closely linked to Reddit's bans. Many of these banned communities exemplify ``e-extremism,'' decentralized networks of online actors spreading conspiracy theories and extremist ideologies through loosely connected communities that blur mainstream-fringe boundaries \citep{zhong2025subdued}. The June 2015 harassment policy ban was followed by an influx that crashed Voat's servers \citep{Newell2016UserMigration}; subsequent bans of Pizzagate (2016), QAnon communities (2018) \citep{Papasavva2021Qoincidence, papasavva2023waitin}, and The\_Donald (2020) \citep{cima2024great, ribeiro2021platform} were likewise followed by measurable migration waves.

Prior research focuses on the \textit{source} platform (does Reddit improve?) and the \textit{banned} communities (do displaced users radicalize?) \citep{ali2021understanding, ribeiro2021platform}. A third perspective remains underexamined: the impact on the \textit{receiving} platform's existing ecosystem. This gap is consequential. As \citet{monti2023online} note, FatPeopleHate was banned in 2015 ``when Voat was a very small platform without an active community,'' while QAnon communities were banned in 2018 when ``several QAnon-related communities existed already on Voat'' and ``the user base in Voat considerably grew.'' The state of the receiving platform, including its existing communities, their structure, and their norms, may shape migration outcomes as much as the properties of displaced users themselves. Yet this dimension remains largely unexplored. When toxic communities migrate, they do not arrive in a vacuum. Alternative platforms host pre-existing generalist communities (such as /v/gaming, /v/funny, /v/technology), populated by users who joined for reasons unrelated to ban-driven controversies. What happens to these communities when they share platform space with displaced users?

Understanding these dynamics requires examining both the \textit{outcome} of structural transformation and the \textit{mechanisms} through which it occurs. Prior work on individual newcomers finds rapid norm adaptation: newcomers adjust their behavior to match existing community toxicity levels through observation before extensive engagement \citep{rajadesingan2020quick}. However, mass migration following ban events presents a qualitatively different scenario: when newcomers arrive in sufficient volume, they may transform existing norms rather than adapt to them.

This study addresses two research questions. First, we ask how post-ban arrivals reshape generalist community structure on receiving platforms (\textbf{RQ1}). We examine whether structural transformation follows more complex dynamics than the integration-or-isolation binary assumed by prior migration studies.

Second, we investigate the mechanisms of community transformation (\textbf{RQ2}). Do post-ban arrivals capture hub positions and impose new norms top-down? Or does transformation operate through peripheral dynamics, that is, sheer volume of newcomer participation that bypasses rather than displaces existing structures?

\section{Data and Methods}

\subsection{Dataset}

In this study, we analyze six matched pairs of Voat and Reddit communities: three media-sharing (/v/pics, /v/gifs, /v/videos) and three topic-oriented (/v/gaming, /v/technology, /v/funny), which represent core generalist content expected on any Reddit-like platform, independent of controversial migrations. 

These data were collected from the Multi-Platform Aggregated Dataset of Online Communities (MADOC) \citep{dankulov2025madoc}, which provides comprehensive longitudinal data for both Reddit and Voat. %introduce what is subverse and subreddit
Both platforms organize users into topic-based communities (subreddits on Reddit, subverses on Voat) where users submit content and engage through comments and voting. 
For Voat, the dataset spans the platform's full lifetime from November 2013 (pre-launch development period; official launch April 2014) through shutdown in December 2020, encompassing over 2.3 million submissions and 16 million comments across 7,095 subverses \citep{mekacher2022voat}. For Reddit, we analyze data from January 2014 through January 2021, encompassing over 227 million posts and comments across the six matched communities. The scale difference between platforms is substantial: Reddit's /r/funny alone averages over 250,000 monthly active users, while Voat's /v/funny accumulated approximately 43,000 total users across its entire seven-year lifetime. This asymmetry means that even small fractions of displaced Reddit users can constitute large influxes relative to Voat's existing population. We define monthly active users as distinct users with at least one post or comment in that month; when reporting reputation-active users we restrict to users with mean monthly reputation $> 1$.
To select communities for this study, we began by identifying the top 50 Voat subverses by overall activity. From this set, we excluded communities primarily dedicated to hate speech (e.g., /v/fatpeoplehate), conspiracy theories (e.g., /v/QRV, /v/pizzagate, /v/GreatAwakening), or political communities (e.g., /v/politics, /v/TheDonald), as well as Voat-specific spaces without Reddit equivalents (e.g., /v/AskVoat, /v/MeanwhileOnReddit). While prior research has extensively documented toxic and extremist communities on alternative platforms \citep{Papasavva2021Qoincidence,Russo2023Spillover,papasavva2024comprehensive}, generalist communities (which constitute the majority of platform activity and user base) remain comparatively understudied, despite being exposed to the same migration pressures. This selection yielded six largest Voat subverses---/v/funny, /v/technology, /v/videos, /v/gaming, /v/pics, and /v/gifs---and their direct Reddit counterparts. On Reddit, these are among the platform's most prominent communities: /r/funny and /r/pics consistently rank in the top 10 subreddits by subscriber count, while /r/gaming, /r/videos, /r/technology, and /r/gifs rank in the top 50. Selected communities span entertainment content (/v/funny, /v/gifs, /v/videos), visual media (/v/pics), and hobby/technology discussion (/v/gaming, /v/technology), providing broad coverage of generalist Reddit-style content categories and allowing us to examine how migration dynamics associated with major Reddit bans affect non-target communities. 

Table~\ref{tab:focus_communities} presents the scale of these communities on Voat. Combined, they represent approximately 170,000 unique users---36\% of all generalist community participants and 4.3\% of all Voat users---and nearly 1.5 million posts and comments. Table~\ref{tab:reddit_communities} presents the corresponding Reddit statistics, illustrating the two-orders-of-magnitude scale difference between platforms.

\begin{table}[h]
\centering
\caption{Focus community statistics on Voat (2013--2020) including rank (position among all Voat subverses by total activity), number of users, total activity (posts and comments), and community category.}
\label{tab:focus_communities}
\begin{tabular}{lrrrc}
\toprule
\textbf{Community} & \textbf{Rank} & \textbf{Users} & \textbf{Activity} & \textbf{Category} \\
\midrule
/v/funny & 10 & 43,253 & 507,496 & Entertainment \\
/v/technology & 12 & 36,524 & 253,852 & Topic-oriented \\
/v/videos & 14 & 28,562 & 248,906 & Media-sharing \\
/v/gaming & 16 & 26,411 & 214,552 & Topic-oriented \\
/v/pics & 17 & 20,821 & 136,195 & Media-sharing \\
/v/gifs & 24 & 14,586 & 78,381 & Media-sharing \\
\midrule
\textit{Combined} & --- & \textit{169,957} & \textit{1,439,382} & --- \\
\bottomrule
\end{tabular}
\end{table}

\begin{table}[h]
\centering
\caption{Focus community statistics on Reddit (2014--2021). Mean monthly active users and total activity (posts and comments) across the study period. Scale ratio indicates how many times larger Reddit activity is compared to Voat.}
\label{tab:reddit_communities}
\begin{tabular}{lrrrc}
\toprule
\textbf{Community} & \textbf{Mean Users/Mo} & \textbf{Total Activity} & \textbf{Scale Ratio} & \textbf{Category} \\
\midrule
/r/funny & 250,805 & 61,861,343 & 122$\times$ & Entertainment \\
/r/pics & 223,035 & 51,917,058 & 381$\times$ & Media-sharing \\
/r/gaming & 178,217 & 44,274,130 & 206$\times$ & Topic-oriented \\
/r/videos & 151,805 & 36,776,651 & 148$\times$ & Media-sharing \\
/r/gifs & 114,570 & 21,090,865 & 269$\times$ & Media-sharing \\
/r/technology & 50,313 & 11,766,134 & 46$\times$ & Topic-oriented \\
\midrule
\textit{Combined} & --- & \textit{227,686,181} & \textit{158$\times$} & --- \\
\bottomrule
\end{tabular}
\end{table}

\subsection{Key Deplatforming Events}

We structure analysis around four canonical Reddit ban events:

\begin{itemize}
    \item \textbf{Event A} (June 10, 2015): FatPeopleHate (FPH) Ban---Reddit's first major mass ban under its harassment policy \citep{chandrasekharan2017cant, Newell2016UserMigration}.
    \item \textbf{Event B} (November 23, 2016): Pizzagate (PG) Ban---removal of a conspiracy theory community \citep{mekacher2022voat}.
    \item \textbf{Event C} (September 12, 2018): GreatAwakening (GA) QAnon Ban---coincided with the largest single-day registration on Voat since 2015 \citep{mekacher2022voat, Papasavva2021Qoincidence}.
    \item \textbf{Event D} (June 29, 2020): The\_Donald (TD) Ban---removal of Reddit's largest pro-Trump community \citep{ribeiro2021platform, Russo2023Spillover}.
\end{itemize}

These events represent four largest documented Reddit bans that induced user migration to Voat \citep{mekacher2022voat, papasavva2023waitin}.

\subsection{Toxicity Detection}

Toxic speech encompasses language that is harmful, abusive, hateful, or intended to harass or demean individuals or groups. While explicit toxicity (direct slurs and overt hate speech) is relatively straightforward to detect using keyword-based methods, implicit toxicity---which relies on coded language, subtle dehumanization, and stereotyping---poses greater challenges for automated detection. Implicit toxic content is particularly prevalent in online communities that have developed strategies to evade content moderation while maintaining hostile discourse norms.

We measure toxicity using RoBERTa ToxiGen \citep{hartvigsen2022toxigen}, a RoBERTa-based classifier specifically trained to detect implicit hate speech and adversarial toxic content. Unlike keyword-based approaches that primarily capture explicit toxicity, ToxiGen is designed to identify subtle and coded toxic language, making it well-suited for analyzing communities where discourse norms may have evolved to circumvent traditional moderation techniques. For a given text, the model determines a toxicity probability score in the interval $[0,1]$, where $0$ indicates non-toxic text and $1$ indicates toxic, hateful, or abusive language. We compute a toxicity score for each comment and post separately.

To construct monthly community series, we first compute per-user monthly mean toxicity across all posts and comments in that month. We then aggregate to the community-month level by taking the unweighted mean of these user-month values, so each active user contributes equally regardless of posting volume. For cross-platform summary statistics (Table~\ref{tab:toxicity_comparison}), we compute active-user--weighted averages across months within the shared 2014--2020 window.

\subsection{Sentiment Analysis}

We additionally measure sentiment using VADER \citep{hutto2014vader}, a rule-based sentiment analysis tool for social media text. Since sentiment temporal trends are qualitatively similar across communities and do not reveal patterns distinct from toxicity, we present sentiment methodology and results in Appendix~\ref{sec:sentiment}.

\subsection{User Cohort Partitioning}

To distinguish migration effects from organic evolution, we partition users into cohorts relative to the ban events using retrospective labeling. For any given month, we define event periods bounded by consecutive ban events. Users whose first activity in the community occurred in any month within the current event period are labeled \textbf{Newcomers} for all months in that period. Users whose first activity occurred before the most recent ban event (relative to the current month) are labeled \textbf{Existing}. We use post-ban arrival cohorts as a proxy for migrants from banned Reddit communities, acknowledging that this approximation includes both displaced users and those who discovered Voat independently during these periods. Importantly, these labels are dynamic: a user who is a Newcomer in one period transitions to Existing in subsequent periods. For example, consider September 2016 (between the June 2015 FPH ban and the November 2016 PG ban). Any user whose first activity was after June 2015 and before November 2016 is labeled Newcomer in September 2016. Any user whose first activity was before June 2015 is labeled Existing. The same user who was a Newcomer in September 2016 would be labeled Existing in any month after the PG ban.

\subsection{Network Metrics}

To characterize community structure, we construct monthly user-interaction networks where nodes represent users and edges represent reply interactions between them. Since we do not have access to second-level comments, we consider only comment-to-post replies. Networks are undirected and unweighted, providing a parsimonious representation of inter-user interaction structure.

For each monthly network we compute:

\begin{itemize}

\item \textbf{Degree distribution}: We compute the degree distribution of the network, which is the distribution of the number of edges each node has.

\item \textbf{Degree Gini}: Measures inequality in the degree distribution using the Gini coefficient \citep{gastwirth1972estimation}, ranging from 0 (perfect equality) to 1 (maximum inequality). High values indicate pronounced hub structure where a few users dominate connectivity; low values indicate egalitarian distribution. We define hubs as the top decile of the degree distribution.

\item \textbf{Degree Assortativity}: Measures the tendency of nodes to connect with others of similar degree, computed as the Pearson correlation coefficient between the degrees of connected nodes \citep{newman2002assortative}. Values range from $-1$ (disassortative: high-degree nodes connect to low-degree nodes) to $+1$ (assortative: high-degree nodes connect to other high-degree nodes). Negative values indicate hub-spoke structures where central users engage peripheral members.

\item \textbf{Krackhardt's E-I Index}: Measures cohort (Newcomers vs Existing users) homophily as $(E - I) / (E + I)$, where $E$ represent the number of edges with outside one's cohort and $I$ represents the number of edges within one's cohort \citep{krackhardt1988informal}. Values range from $-1$ to $+1$, where negative values indicate inward orientation (within-cohort segregation: users interact primarily with same-cohort members) and positive values indicate outward orientation (cross-cohort integration: users interact primarily with other-cohort members).

\item \textbf{Newcomer Hub Rate}: Proportion of newcomers whose degree exceeds the existing-user hub threshold (top decile of existing users' degree distribution), measuring whether post-ban arrivals achieve structural centrality within the established elite.

\item \textbf{Degree Share / Population Share}: Ratio of newcomers' degree distribution to their population share, measuring whether newcomers are over-represented (ratio > 1) or under-represented (ratio < 1) in the network.

\end{itemize}

\subsection{Dynamic Reputation}

We employ the Dynamic Interaction Based Reputation Model (DIBRM) to quantify social trust evolution \citep{vranic2023sustainability}. Following \citet{vranic2023sustainability}, we compute per-user reputation from time-ordered interaction events (posts and comments) with exponential inactivity decay. After each interaction, reputation is updated as the sum of a decayed previous value (forgetting factor $\beta$) and a streak-dependent increment $I(A)=I_b + I_b\alpha\left(1-\frac{1}{A+1}\right)$, where $A$ counts consecutive interactions since the last inactivity gap. We use $I_b=1$, $\alpha=2$, and $\beta=0.96$. Reputation grows through the cumulative effects of frequent, sequential interactions and decays via the forgetting factor that penalizes inactivity, reflecting the asymmetric nature of trust (easier to lose than to gain). Prior research demonstrates that stable, high-reputation ``cores'' are crucial for community sustainability \citep{vranic2023sustainability}.

For community-level summaries, we report mean reputation over active users (reputation $\ge 1$), computed by averaging daily means within each month. Core/periphery reputation summaries are obtained by joining user-day reputations to monthly core/periphery labels. In \citet{vranic2023sustainability}, we provide a detailed analysis of parameter choices for measuring reputation evolution in Stack Exchange communities. We argue that Stack Exchange, Reddit, and Voat have similar structural organization and dynamics; therefore, we use the same parameter values here.

\subsection{Regime Breakpoint Validation}

If major ban events coincide with structural regime changes in Voat's generalist communities, we should observe discontinuities in community metrics aligned with those events. To test this, we apply a single-break segmented regression to each Voat monthly time series. For a given metric $y_t$ (toxicity, E-I index, existing-user degree Gini, degree assortativity, and mean reputation), we fit a piecewise linear model with one breakpoint $\tau$, selecting the $\tau$ that minimizes the sum of squared residuals (SSE) while requiring at least six months of data on each side. Break strength is summarized by the SSE ratio (segmented model SSE divided by single-line SSE), where lower values indicate stronger breaks. We assess breakpoint stability via residual bootstrap over 200 iterations, yielding 5th--95th percentile confidence intervals and reporting the proportion of bootstrap samples where the estimated break falls within $\pm$3 months of the point estimate. We run this validation per community and on the global mean series.
Because monthly series exhibit temporal dependence, the bootstrap intervals should be interpreted as descriptive stability checks rather than formal time-series confidence intervals (they do not explicitly model autocorrelation).

\section{Results}

\subsection{Platform Dynamics}

\begin{figure}[H]
\centering
\includegraphics[width=0.7\textwidth]{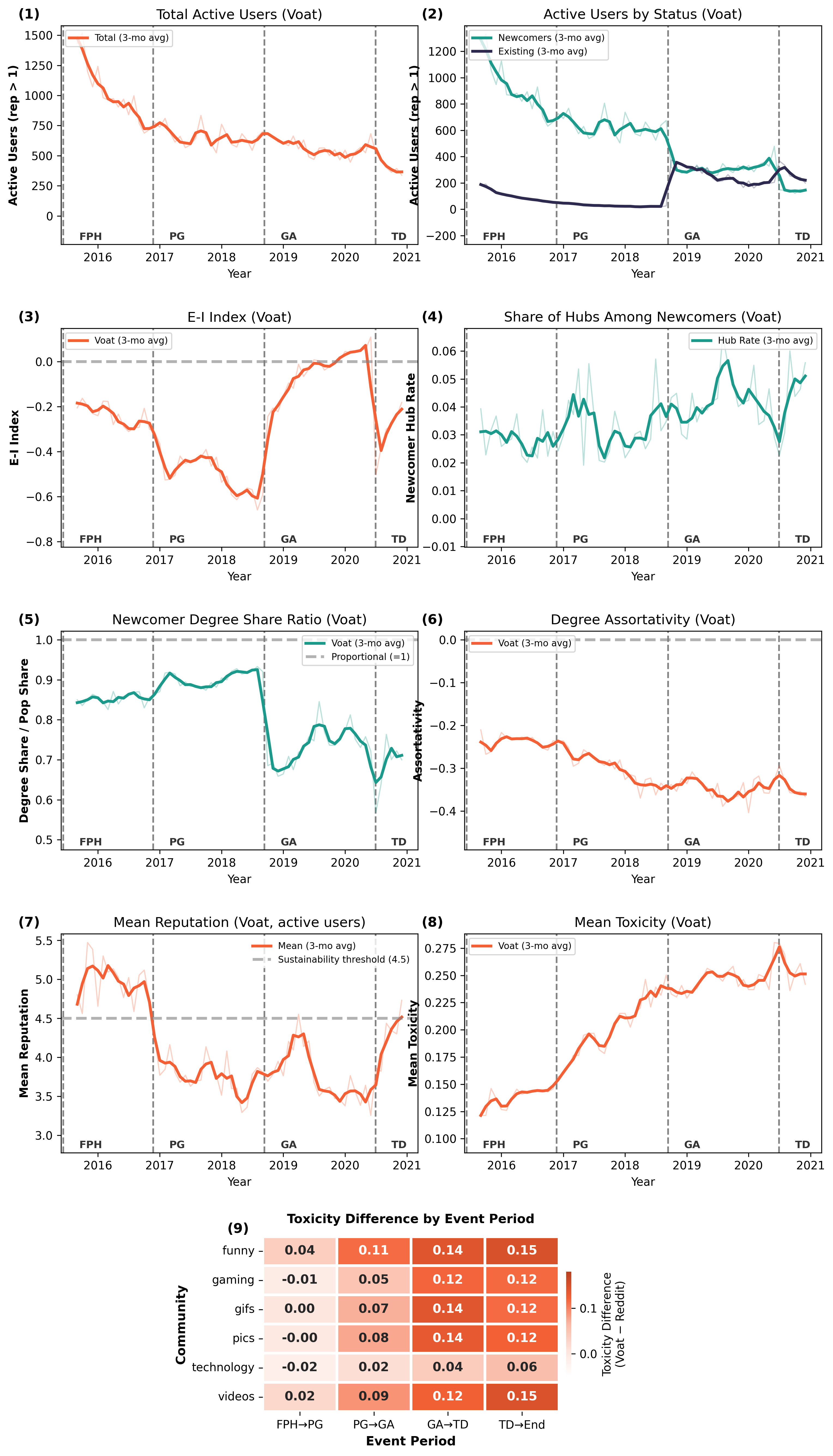}
\caption{Platform-level dynamics on Voat across generalist communities. Vertical dashed lines indicate ban events (FPH: fatpeoplehate, PG: Pizzagate, GA: GreatAwakening, TD: The\_Donald). Panels show: (1) total active users with reputation $> 1$, (2) active users split by newcomer/existing status, (3) E-I Index measuring cohort segregation, (4) newcomer hub rate, (5) degree share ratio, (6) degree assortativity, (7) mean reputation with 4.5 sustainability threshold (dashed line), (8) mean toxicity on Voat, and (9) toxicity differential heatmap by event period.}
\label{fig:global_summary}
\end{figure}

Figure~\ref{fig:global_summary} shows Voat's aggregated dynamics across the six generalist communities included in our analysis (\textit{funny}, \textit{gaming}, \textit{gifs}, \textit{pics}, \textit{technology}, \textit{videos}). Global time series are computed as weighted averages of community-level monthly measures, with weights proportional to each community's user count (or users with reputation $\ge 1$ for reputation metrics). A three-month centered rolling average is applied to all time series for smoothing. Because aggregation is at the community-month level, users active in multiple communities contribute to each community's weight; the global series should be interpreted as activity-weighted community averages rather than unique-user averages. Together, the panels reveal a structural transformation of Voat that aligns with four ban events (FPH, PG, GA, TD), indicated as vertical lines.

The total number of active users with reputation $> 1$ (panel 1) peaked at 11{,}267 immediately after the FPH ban (July 2015) and then declined to a steady state of 2{,}000--3{,}000. Panel 2 splits these active users by cohort status; sustained activity (reputation $> 1$) peaked at $\approx 1{,}500$ in late 2015 and stabilized around 340--760 in 2018--2020. Existing users (blue line) declined steadily from $\approx 800$ in 2015 to $\approx 200$ by 2020, forming a shrinking but persistent core, while newcomers (orange line) dominated numerically, comprising 60--80\% of active users in most periods despite their transient status.

The E-I Index (panel 3), measuring cross-cohort interaction between Newcomers and Existing users, exhibits three distinct phases. After the FPH ban (June 2015), the index started at $-0.35$, indicating moderate cohort insularity---users preferentially interacted within their own cohort. It then collapsed steadily to $-0.66$ by August 2018, reflecting extreme within-cohort concentration where Newcomers and Existing users rarely engaged each other. Following the GreatAwakening ban (September 2018), the index rebounded sharply, reaching approximately 0 by mid-2019 and $+0.05$ by early 2020---the only period of net cross-cohort interaction, suggesting temporary integration between cohorts. The TD ban (June 2020) was followed by a secondary collapse to $-0.50$, indicating renewed cohort segregation as post-ban arrivals formed insular interaction patterns.

Despite their numerical dominance, newcomers remained structurally marginalized: the newcomer hub rate (panel 4) remained low at 2--6\% throughout (median 3.3\%), indicating that few newcomers exceeded the existing-user top-decile degree threshold. The degree share ratio (panel 5)---defined as newcomers' share of total network degree divided by their share of total users---ranged from $0.84$--$0.92$ in early periods (2015--2018), then dropped sharply after the GA ban to $0.66$--$0.76$ (2018--2020), with a deeper dip to $0.57$--$0.72$ following the TD ban. A ratio of 1.0 (horizontal dashed line) indicates connections proportional to population share; values below 1.0 indicate newcomers systematically received fewer connections than their population share would predict. This finding is robust to alternative hub thresholds (5\%, 20\%; see Appendix Table~\ref{tab:hub_sensitivity}).
Network structure became increasingly disassortative over time (panel 6), with degree assortativity shifting from $-0.14$ (July 2015) to $-0.37$ (late 2018--2020), indicating that high-degree users increasingly interacted with low-degree users, which is consistent with core members engaging newcomers rather than forming insular hubs.

Mean reputation (panel 7) declined from $\approx 4.5$--5.5 in 2015--2016 to $\approx 3.5$--4.0 by 2018--2020, reflecting either declining engagement quality or the dilution effect of successive low-reputation newcomer cohorts. The horizontal dashed line at 4.5 marks the sustainability threshold identified by \citet{vranic2023sustainability}: failed Stack Exchange communities exhibited mean reputation declining below this level in their final days before collapse, a threshold generalist communities on Voat crossed by 2018, suggesting they may have entered a similar trust-deficient state.

Mean toxicity (panel 8) rose steadily from $\approx 0.12$ in late 2015 to over $0.24$ by mid-2018, effectively doubling within three years. The trajectory plateaued around $0.25$ thereafter, suggesting a new baseline toxicity level stabilized by successive migration waves. Table~\ref{tab:toxicity_comparison} quantifies these cross-platform differences: Voat's generalist communities exhibit consistently higher toxicity than their Reddit counterparts, with ratios ranging from 1.28$\times$ (\textit{gifs}) to 1.81$\times$ (\textit{funny}). The exception is \textit{technology}, which maintains comparable levels (1.02$\times$), possibly reflecting topic-focused discussion that provides less opportunity for interpersonal conflict.

\begin{table}[h]
\centering
\caption{Active-user--weighted mean toxicity scores for generalist communities on Reddit vs. Voat (2014--2020, shared window).}
\label{tab:toxicity_comparison}
\begin{tabular}{lcccc}
\toprule
\textbf{Community} & \textbf{Reddit} & \textbf{Voat} & \textbf{Ratio (Voat/Reddit)} & \textbf{Difference (Voat-Reddit)} \\
\midrule
funny & 0.127 & 0.229 & 1.81$\times$ & +0.102 \\
gaming & 0.093 & 0.128 & 1.37$\times$ & +0.035 \\
gifs & 0.137 & 0.176 & 1.28$\times$ & +0.039 \\
pics & 0.125 & 0.186 & 1.49$\times$ & +0.061 \\
videos & 0.138 & 0.209 & 1.51$\times$ & +0.071 \\
technology & 0.133 & 0.136 & 1.02$\times$ & +0.003 \\
\bottomrule
\end{tabular}
\end{table}

The heatmap (panel 9) quantifies the difference in toxicity level between Voat and Reddit for intervals between successive ban events. During the A--B interval (FPH to Pizzagate), differences were modest ($<0.04$). The gap widened progressively through subsequent periods: B--C showed differences of $0.02$--$0.11$, C--D reached $0.04$--$0.14$, and D--end peaked at $0.06$--$0.15$. The \textit{funny} and \textit{videos} communities exhibited the largest divergence ($>0.14$), while \textit{technology} remained most similar to Reddit with differentials of $\approx 0.05$.

\subsection{Case Study: /v/funny}

The global dynamics in Figure~\ref{fig:global_summary} suggest a trajectory of decline: mean reputation falling below the sustainability threshold identified by \citet{vranic2023sustainability}, consistent with failed communities in terminal decay. Yet not all generalist communities followed this path. We turn to /v/funny, the largest generalist community on Voat (43,253 users, 507,496 posts), which presents a counterpoint: a community that not only survived the migration waves but achieved structural stability.

Figure~\ref{fig:funny_overview} illustrates these dynamics. Among all communities studied (see Appendix~\ref{sec:appendix_community_dynamics} for more details on other communities), /v/funny exhibits the highest toxicity ratio relative to its Reddit counterpart (1.81$\times$) and the largest absolute toxicity differential (+0.102). Rather than indicating failure, however, these metrics reflect successful adaptation to a transformed norm environment.

The June 2015 FPH ban triggered immediate transformation. Active users surged on Voat following each ban event (panel 5), while Reddit's /r/funny remained stable and unaffected (panel 4). Unlike other generalist communities where such influx proved destabilizing, /v/funny absorbed newcomers into sustained growth.

The most striking pattern is the toxicity divergence (panel 1). Mean toxicity rose from $\sim$0.10 (early 2015) to over 0.25 by 2018, more than doubling within three years, while Reddit's /r/funny remained flat (mean: 0.127). This divergence widened with each subsequent ban event and persisted through platform shutdown---migration effects compounded rather than dissipated over time.

Crucially, this toxic transformation did not produce community collapse. Mean reputation (panel 2) stabilized following initial turbulence and approached levels comparable to its Reddit counterpart by 2020, the hallmark of community health in the \citet{vranic2023sustainability} framework. The remaining panels (3, 6, 7) show network structure metrics consistent with this interpretation: lower connectivity reflecting smaller community size, and negative assortativity indicating hub-mediated interaction patterns typical of online communities.

Crucially, the reputation stabilization reveals that the toxic transformation was not merely disruptive but \textit{generative}: the community achieved sustainability. Because reputation in our framework is activity-based, stable levels indicate that users placed sufficient trust in the community to sustain their participation, a nontrivial commitment given that the majority of users on social platforms remain passive consumers rather than active contributors. This suggests that toxic norms became self-reinforcing through normal community dynamics.

\begin{figure}[H]
\centering
\includegraphics[width=0.6\textwidth]{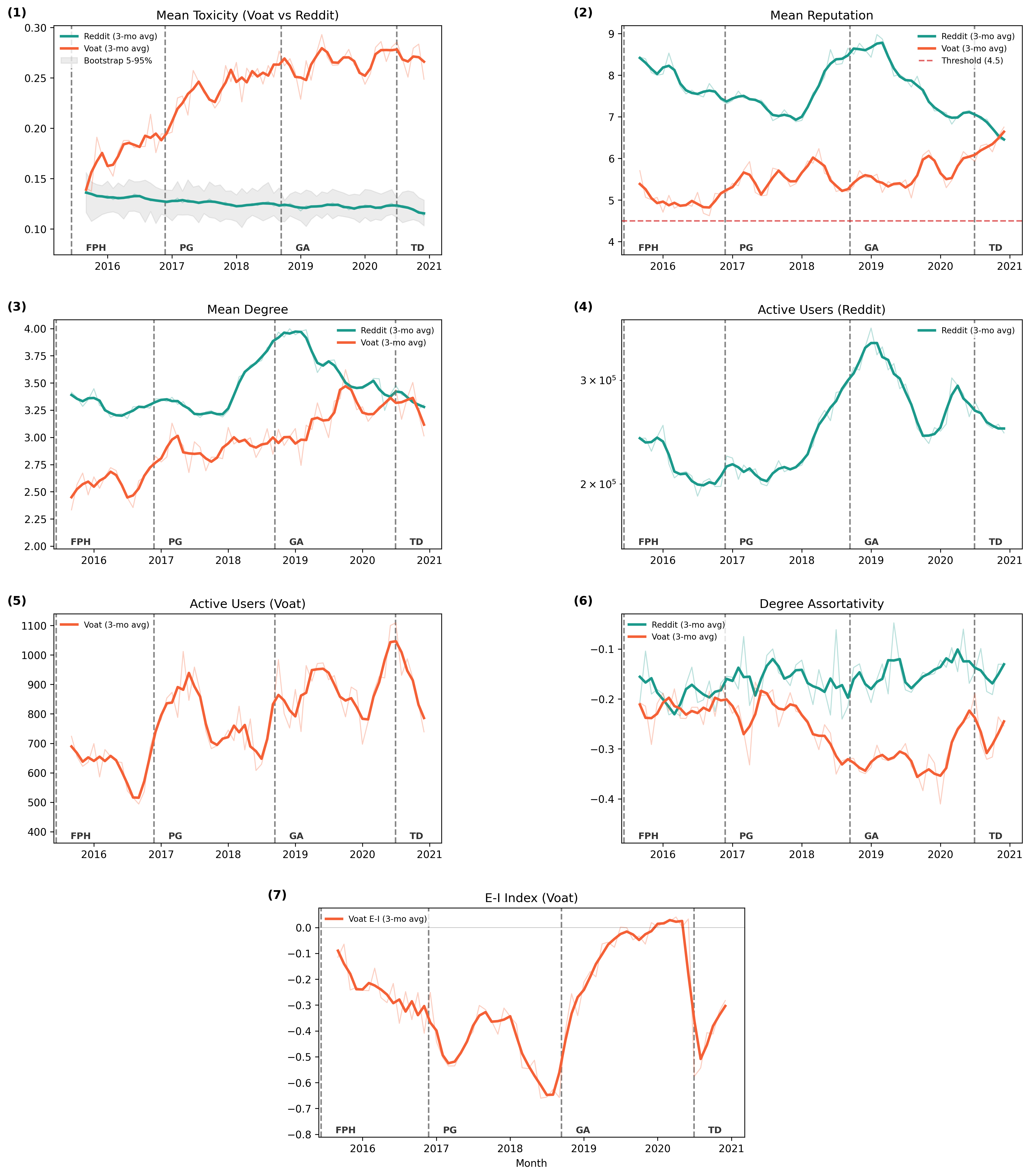}
\caption{Dynamics of /v/funny (Voat) vs. /r/funny (Reddit). The progressive toxicity increase following Event A demonstrates the Hostile Takeover regime; the plateau after Event C marks the transition to Toxic Equilibrium.}
\label{fig:funny_overview}
\end{figure}

\subsection{Breakpoint Validation of Regime Change}

We validate the two-regime interpretation using the single-break segmented regression described in Methods. Table~\ref{tab:breakpoint_summary} summarizes the results across metrics.

\begin{table}[h]
\centering
\caption{Breakpoint analysis summary by metric. ``Near GA'' counts series with breaks within $\pm$3 months of September 2018. Lower SSE ratios indicate stronger breaks.}
\label{tab:breakpoint_summary}
\small
\begin{tabular}{lccp{5cm}}
\toprule
\textbf{Metric} & \textbf{Near GA} & \textbf{SSE Ratio} & \textbf{Break Months Observed} \\
\midrule
E-I Index & 5/7 & 0.36 & 2017-02, 2018-10, 2019-01 \\
Existing Gini & 0/7 & 0.18 & 2015-07 \\
Degree Assortativity & 0/7 & 0.46 & 2015-01 to 2015-09 \\
Mean Toxicity & 0/7 & 0.60 & 2015-02 to 2020-03 \\
Mean Reputation & 1/7 & 0.64 & 2014-11, 2015-01, 2015-08, 2016-05, 2016-06, 2016-10, 2018-07 \\
\bottomrule
\end{tabular}
\end{table}

The E-I index shows robust breakpoints aligned with the GA ban: five of seven series (four communities plus global) exhibit breaks within $\pm$3 months of September 2018, with a median SSE ratio of 0.36 indicating strong model fit improvement. Four communities (\textit{gaming}, \textit{pics}, \textit{technology}, \textit{videos}) and the global aggregate show breakpoints in October 2018 with bootstrap confidence intervals concentrated entirely on that month (stability = 1.0). /v/funny shows a closely following breakpoint in January 2019, while /v/gifs exhibits a weaker earlier break (February 2017, SSE ratio = 0.52). The regime pattern persists under an alternative rolling-tenure newcomer definition (Appendix Table~\ref{tab:ei_rolling}).

In contrast, structural hierarchy metrics show breakpoints aligned with the earlier FPH migration. Existing-user degree Gini exhibits the strongest breaks in the dataset: all seven series break in July 2015 with median SSE ratio of 0.18 and perfect bootstrap stability (90\% CI collapsed to a single month: July 2015 in all series, indicating zero variance across bootstrap iterations). Degree assortativity similarly concentrates in early-to-mid 2015, with median SSE ratio of 0.46. These patterns indicate that hierarchy and mixing structure were reconfigured immediately following the first major migration wave.

Mean toxicity displays dispersed breakpoints across communities (2015--2020) with median SSE ratio of 0.60, consistent with a gradual rise and plateau rather than an abrupt shift at any single event. Mean reputation likewise shows heterogeneous breakpoints, with only /v/technology exhibiting a near-GA break (July 2018); the remaining series break earlier and do not align with specific ban events. Full breakpoint estimates appear in Table~\ref{tab:breakpoints_full} (Appendix).

\section{Discussion}

\subsection{Two Regimes of Migration Impact}

The results reveal a critical inflection point: the September 2018 QAnon ban coincides with a structural discontinuity in how generalist communities responded to migration. Before this point, we observe a consistent pattern of deterioration: E-I index collapsing from $-0.35$ to $-0.66$, toxicity escalating from $0.12$ to $0.25$, and mean reputation declining below the sustainability threshold identified by \citet{vranic2023sustainability} for other social platforms. In the period after the QAnon ban, the trajectory reverses: E-I index recovers to $+0.05$, toxicity plateaus rather than continues rising, and, in the /v/funny case, reputation stabilizes at sustainable levels.

What accounts for this reversal? Crucially, the newcomer hub rate remained constant at 2--6\% (median 3.3\%) throughout both periods. Fewer than 5\% of newcomers achieved degree centrality exceeding the existing-user hub threshold in most months, yet community transformation proceeded regardless. This stability rules out hub capture as the mechanism of change and directs attention elsewhere: to how post-ban arrivals interacted with existing structure, and how that structure itself evolved. The divergent trajectories before this event suggest two qualitatively different dynamics at work.

\subsubsection{Regime 1: Hostile Takeover Through Volume (2015--2018)}

The period following early ban waves exhibits a characteristic signature: E-I index collapse ($-0.35$ to $-0.66$), rising toxicity ($0.12$ to $0.25$), stable existing-user Gini ($\approx 0.70$), and declining reputation.

This pattern reflects a \textit{hostile takeover through volume, not hub capture}. Post-ban arrivals participated actively at the periphery with high motivation. The newcomer hub rate remained at 2--6\% (median 3.3\%)---they did not displace existing hubs. Instead, they \textit{ignored} them.

The collapsing E-I index reflects this bypass. Newcomers interacted primarily with other newcomers, forming parallel social structures alongside the existing hub hierarchy. The original community core remained structurally intact but became increasingly irrelevant as the newcomer-dominated parallel network grew around it.

Reputation decline accompanied structural marginalization. Trust-building requires engagement with recognized community members. With newcomers bypassing existing hubs, the reputation system lost coherence.

\subsubsection{Regime 2: Toxic Equilibrium Through Integration (2018--2020)}

The post-QAnon period exhibits a different pattern: E-I index recovery ($-0.66$ to $+0.05$), toxicity plateau (0.25), flattened existing-user Gini ($\approx 0.58$), and reputation stabilization. The /v/funny case study exemplifies this transition: reputation levels stabilized at sustainable values, indicating a structurally viable community rather than one in decline.

This reflects the emergence of a stable \textit{toxic equilibrium}. Post-ban arrivals from QAnon communities concentrated in dedicated conspiracy subverses rather than flooding generalist communities. Those who arrived at /v/funny or /v/gaming were casual, peripheral participants. Crucially, the existing user hierarchy had itself flattened. The pronounced hub structure of Regime 1 gave way to a more egalitarian distribution. Whether through hub departure, reduced activity, or network contraction, the community no longer had a distinct structural elite.

The rising E-I index indicates genuine cross-cohort integration. Peripheral newcomers engaged with the flattened existing population rather than forming parallel structures. But those ``existing'' users were post-ban arrivals from Regime 1 who had established toxic discourse as the norm. Rather than bypassing existing structure, Regime 2 newcomers were absorbed into it.

Independent evidence corroborates this equilibrium. \citet[pg.~5875]{dimarco2024users}, focusing on user behavior on Reddit and Voat in 2019 (middle section of Regime 2), find that on Voat, unlike Reddit, ``most of the subverses have a sublinear behavior, i.e., users tend to stay longer in the same communities.'' This persistence contrasts with Reddit's superlinear decay, where users cycle through communities more rapidly. The Voat pattern suggests that once post-ban arrivals became the community during Regime 1, they persisted—creating the stable conditions we observe in Regime 2. The toxic equilibrium was not merely a transient state but a self-reinforcing configuration where transformed norms reproduced through sustained user participation.

The contrast illuminates two mechanisms. Regime 1 operated through \textit{volume}: newcomers formed parallel structures and overwhelmed existing dynamics through numbers. They did not capture hubs, they made them irrelevant.

Regime 2 operated through \textit{integration}: fewer newcomers arrived, participated peripherally, but were socialized by the now-dominant toxic community. The flattened hierarchy enabled this: newcomers simply joined the community as it was, with no structural elite to bypass.

The stable 2--6\% hub rate (median 3.3\%) across both regimes confirms that neither mechanism required capturing structural centrality. Community transformation occurred at the periphery.

These findings directly address our research questions. Regarding \textbf{RQ1} (how post-ban arrivals reshape generalist community structure), we observe a two-phase transformation: initial parallel structure formation during Regime 1, where newcomers bypassed existing community cores through sheer volume, followed by integration into already-transformed communities during Regime 2, where the flattened hierarchy enabled absorption rather than bypass. The E-I index trajectory---collapsing from $-0.35$ to $-0.66$ (Regime 1) then recovering to $+0.05$ (Regime 2)---quantifies this structural shift from segregation to integration. Regarding \textbf{RQ2} (mechanisms of community transformation), the evidence consistently points to peripheral dynamics rather than hub capture. Newcomer hub rates remained at 2--6\% throughout both regimes, yet toxicity doubled and community norms transformed fundamentally. Transformation operated through volume-driven bypass of existing structures (Regime 1) and subsequent socialization into toxic norms (Regime 2), not through displacement of influential users.

\subsection{The success of /v/funny}

When we look at all six generalist communities, we observe that in regime 2, newcomers engage with them in a more casual, peripheral manner. Toxicity plateaus and interactions between existing users from previous ban waves and newcomers become more heterogeneous. The number of active newcomers and existing users is balanced and stable in the period after the last ban. Voat's lifetime was abruptly cut short by the platform's shutdown in December 2020, but the evidence suggests that the communities were able to sustain themselves through the migration waves and the ban events.

The largest generalist community on Voat, /v/funny, presents a particularly instructive case. Its mean reputation and mean degree levels at the end of the study period are comparable to its Reddit counterpart, suggesting the community achieved sustainability through the migration waves despite becoming substantially more toxic.

\subsection{Implications for platform governance}

Our findings complicate the narrative that deplatforming simply ``cleans'' the internet. While source platforms benefit \citep{chandrasekharan2017cant}, the costs are externalized to receiving platforms—and specifically to their generalist communities, not just dedicated hate spaces. The two-regime pattern suggests a critical window for intervention, assuming the receiving platform would like to preserve founding values and community norms, at least in generalist communities, and not become a dedicated hate space itself.

Regime 1 (hostile takeover) is when lasting transformation occurs: post-ban arrival cohorts form parallel structures, bypass existing community cores, and establish new toxic norms before platform administrators can respond. By Regime 2, transformation is complete: newcomers simply integrate into the already-toxic community. This implies that receiving platforms face a narrow window after major ban events when intervention might prevent permanent norm shifts. The mechanism matters for policy design. Because transformation operated through peripheral volume rather than hub capture, moderation strategies targeting influential users or extreme examples of toxic behavior may have been ineffective. The newcomer hub rate remained low (median 3.3\%), yet /v/funny's toxicity doubled.

This suggests that volume-based interventions such as: rate limiting new accounts, staggered onboarding, or temporary restrictions on newcomer posting, may be more effective than targeting visible bad actors. Finally, path-dependency complicates reversal. Once toxic norms stabilized through sustained participation \citep{dimarco2024users}, the community achieved equilibrium. Voat's generalist spaces never recovered.

This has important implications for deplatforming decisions. Banning communities when alt-platforms are small and unstructured (as with FPH in 2015) may produce greater collateral effects on generalist spaces than banning when dedicated enclaves already exist (as with QAnon in 2018).  Paradoxically, larger platforms are more exposed to the public scrutiny and more likely to be banned themselves by the hosting providers or regulators. This is because they are more visible and more likely to be targeted by the public and the media. Smaller platforms may survive under the radar and continue to operate during the hostile takeover regime. For example, Voat’s CEO asked Voat users to stop threatening people as he had been contacted by a “US agency” in April 2019 \citep{mekacher2022voat} when toxicity has already plateaued and the number of both newcomers and existing users is balanced and stable. This is a clear indication that the platform was under pressure from the authorities and was forced to take action.

\subsection{Limitations}

Several limitations constrain our analysis. First, we cannot establish causation between ban events and community changes. While Voat's dynamics were largely shaped by migration, our results should be interpreted as descriptive rather than causal. Potential confounds include Voat's permissive moderation policies, self-selection of users who chose the platform, and organic platform maturation—factors that could independently drive toxicity increases and might coincide temporally with Reddit's ban schedule.

Second, we cannot track individual users across platforms. Our Newcomer/Existing partition captures timing of first community activity but not user intent or origin. We use post-ban arrival cohorts as a proxy for displaced users, but this approximation necessarily includes both actual migrants from banned Reddit communities and users who discovered Voat independently during these periods. A user appearing after the FPH ban might be a dedicated displaced user, a casual Reddit user, or someone who discovered Voat through other channels. This conflation may overstate migration effects if organic growth coincided with ban events.

Third, our network construction relies on first-level comments only; comment-to-comment threading is unavailable in the dataset. Peer-to-peer discussions within threads remain invisible to our analysis, and the ``parallel structures'' we identify could alternatively reflect thread-level conversational dynamics. Fourth, monthly aggregation may mask critical phase transitions in the immediate post-ban period, when the first days or weeks could differ meaningfully from later patterns. Fifth, while ToxiGen RoBERTa captures implicit hate speech effectively, it may miss Voat-specific coded language and cultural markers. Longitudinal drift in toxic discourse conventions over the seven-year study period may also affect measurement consistency.

Despite these limitations, our study offers distinctive contributions that stem precisely from our methodological choices. The constraints that limit causal inference enabled comprehensive longitudinal coverage spanning Voat's complete lifespan—a scope that would be impossible with richer but more limited data. This breadth reveals the two-regime pattern (hostile takeover followed by toxic equilibrium) that emerges only at multi-year scales and would be invisible in shorter observation windows. Our focus on generalist communities, rather than dedicated hate spaces, illuminates an understudied dimension of deplatforming's collateral effects. The matched cross-platform comparisons with Reddit provide counterfactual baselines demonstrating that observed changes reflect platform-specific dynamics rather than broader temporal trends in online discourse. Multiple complementary metrics—toxicity, network structure, cohort dynamics, and reputation—converge on consistent findings: community transformation occurred through peripheral volume rather than hub capture, with newcomer hub rates remaining low (median 3.3\%) even as toxicity doubled. This convergence across six communities and multiple analytic approaches suggests our core findings are robust to particular measurement choices.

\section{Conclusion}

This study provides the first systematic analysis of how deplatforming affects generalist communities on receiving platforms—communities that were never intended to host displaced users but absorbed them nonetheless. By combining network analysis, toxicity detection, and dynamic reputation modeling across Voat's complete seven-year lifespan, we reveal structural mechanisms of community transformation that would be invisible to shorter-term or single-metric approaches.

Our central finding is the two-regime pattern governing migration impact: \textit{hostile takeover} through peripheral volume (2015--2018), followed by \textit{toxic equilibrium} through integration (2018--2020). Transformation did not require newcomers to capture influential positions—fewer than 5\% of newcomers achieved degree centrality exceeding the existing-user hub threshold in most months throughout both regimes. Instead, post-ban arrival cohorts overwhelmed communities through sheer participation volume, forming parallel structures that bypassed rather than displaced existing hubs. When community hierarchy subsequently flattened, toxic norms reproduced through diffuse peer interaction rather than top-down influence.

The /v/funny case study crystallizes these dynamics: a community that doubled its toxicity relative to Reddit yet achieved sustainable reputation levels by 2018. This was not community decay but community success—organized around transformed norms. The computational social science contribution lies in demonstrating that peripheral dynamics, typically dismissed as noise around influential hubs, can drive fundamental community transformation when volume is sufficient.

These findings carry implications for platform governance. Deplatforming effectively cleans source platforms, but the costs are externalized to receiving ecosystems. The two-regime pattern suggests a critical intervention window: Regime 1 is when lasting damage occurs through parallel structure formation; by Regime 2, transformation is complete and newcomers simply integrate into already-toxic communities. Volume-based interventions (rate limiting, staggered onboarding) may prove more effective than targeting visible bad actors, since transformation operates at the periphery rather than through hub capture.

More broadly, this study demonstrates the value of matched cross-platform comparison for causal reasoning in observational settings. Reddit's stable toxicity levels across the same period provide a counterfactual baseline, strengthening confidence that observed changes on Voat reflect migration dynamics rather than broader temporal trends in online discourse. The convergence of multiple metrics—toxicity, network structure, cohort dynamics, and reputation—on consistent findings across six communities suggests robust underlying mechanisms rather than measurement artifacts.

Voat shut down in December 2020, and its users dispersed to subsequent alt-platforms. Whether the toxic norms acquired through egalitarian diffusion on Voat transferred to Parler, Gab, or other destinations remains an open question for future research. The framework developed here—tracking regime transitions through peripheral dynamics and hierarchy flattening—offers a template for studying these ongoing migrations. As deplatforming continues to reshape the online ecosystem, understanding its full consequences requires attending not only to banned communities and source platforms, but to the generalist spaces that absorb the displaced.

\section*{Data Availability}

All datasets analyzed in this study are available through the Multi-Platform Aggregated Dataset of Online Communities (MADOC) archive at https://zenodo.org/records/14637314. The analysis code and derived results are available at https://github.com/atomashevic/Reddit-Voat-comparison.

\section*{Acknowledgments}

This research was financially supported by the Science Fund of the Republic of Serbia, Prizma program (grant No. 7416). Data processing and analysis were performed on the PARADOX-IV supercomputing facility at the Scientific Computing Laboratory, National Center of Excellence for the Study of Complex Systems, Institute of Physics Belgrade.

Claude 4.5 Sonnet LLM was used in preparation of this manuscript, for table generation, text editing and LaTeX formatting.

\section*{Ethics}

This study analyzes publicly available user-generated content collected in the MADOC dataset and does not involve interaction with users, interventions, or access to private information. The data are de-identified by the dataset providers, and we report results only in aggregate. We did not attempt to re-identify individuals, and no personally identifying information is presented. Under these conditions, the research is exempt from human-subjects review.

\section*{Competing Interests}

The authors declare no competing interests.

\bibliographystyle{unsrtnat}
\bibliography{references}

\newpage
% ============================================================================
% APPENDIX
% ============================================================================
\appendix

\section{Additional Community Dynamics}
\label{sec:appendix_community_dynamics}

This appendix presents overview figures for additional generalist communities. These communities exhibit similar two-regime patterns with varying magnitudes.

\begin{figure}[h]
\centering
\includegraphics[width=0.6\textwidth]{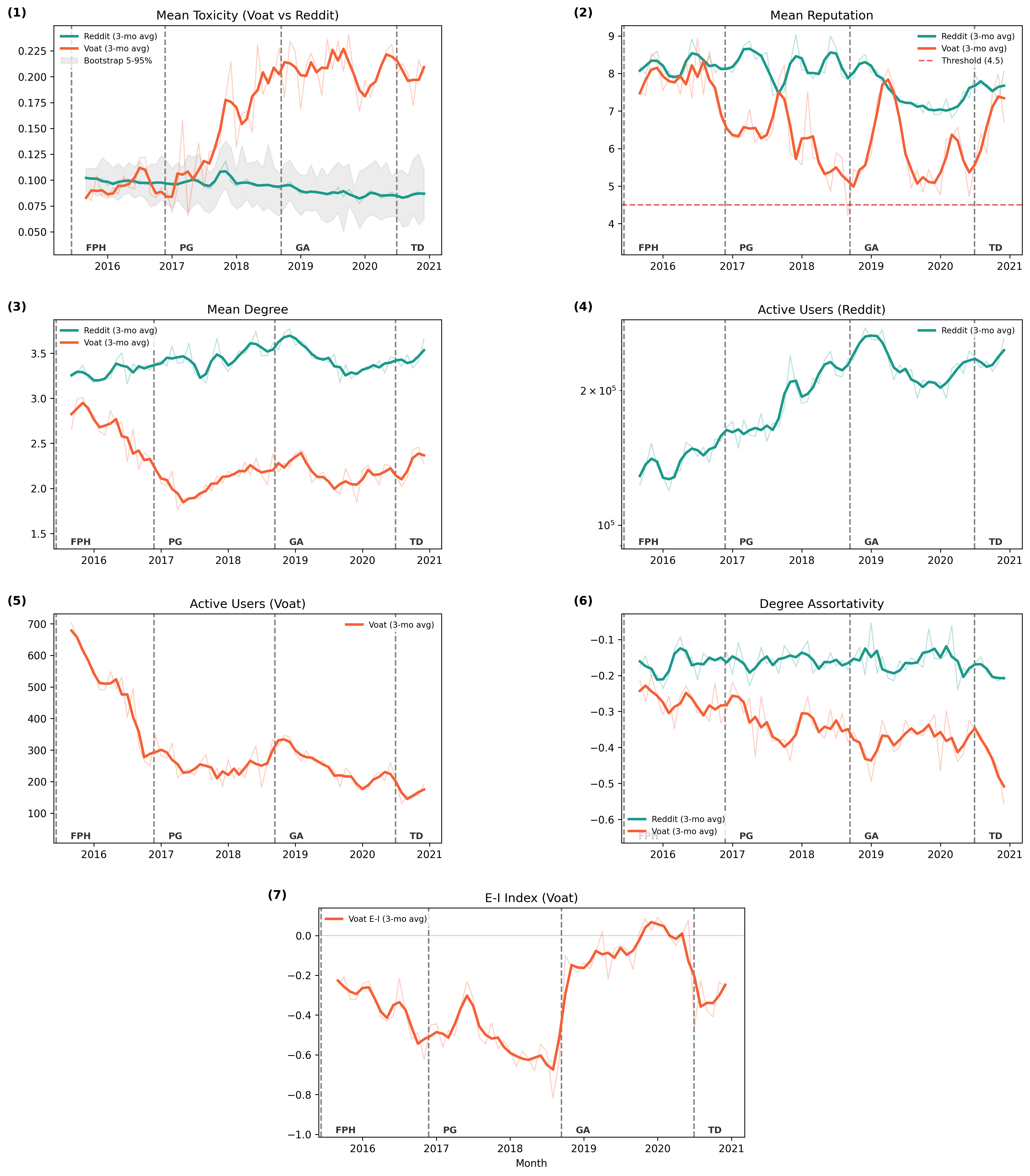}
\caption{Dynamics of /v/gaming vs. /r/gaming. Sharp activity spike following Event A, sustained toxicity elevation through both regimes.}
\label{fig:gaming_overview}
\end{figure}

\begin{figure}[h]
\centering
\includegraphics[width=0.6\textwidth]{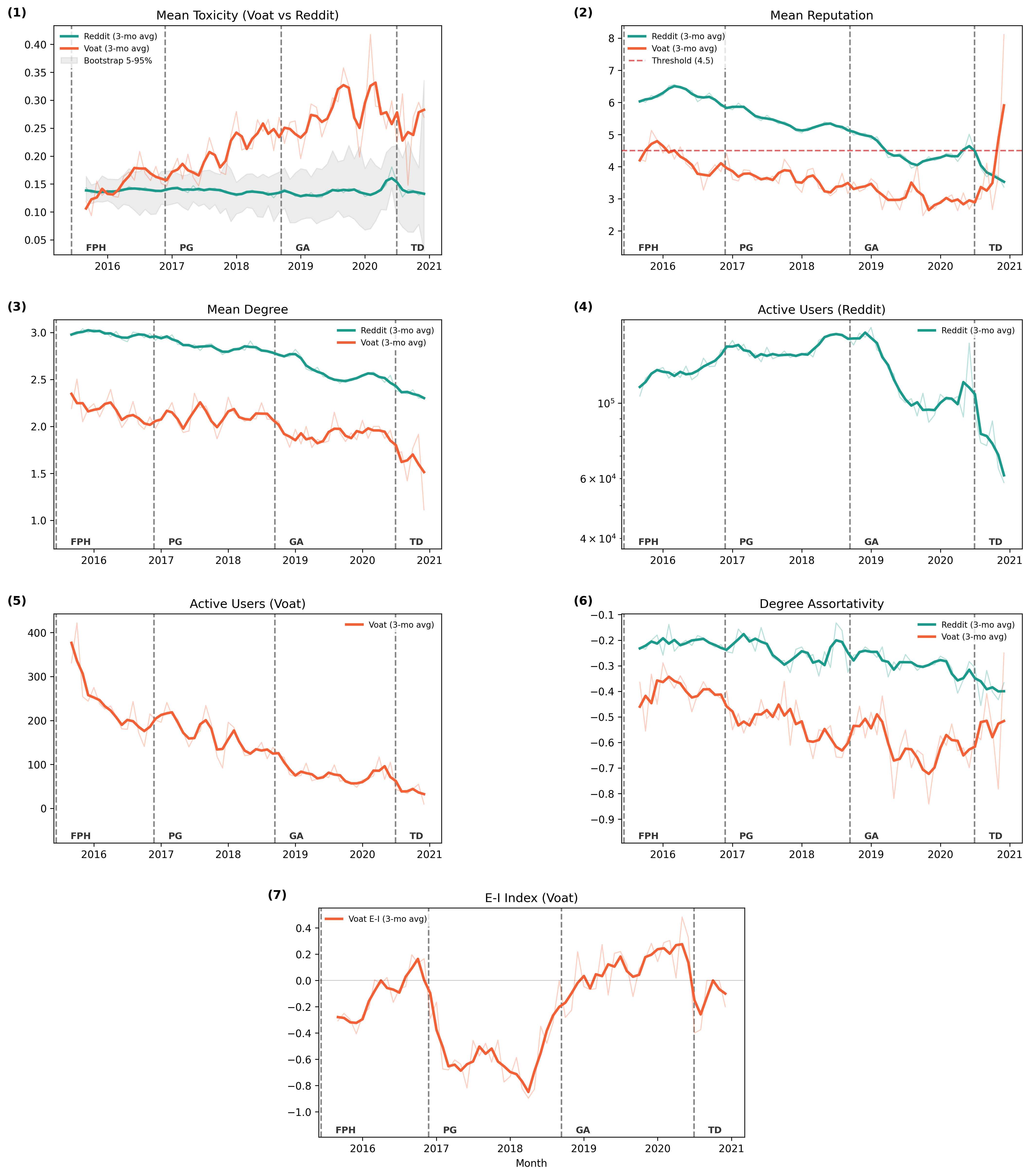}
\caption{Dynamics of /v/gifs vs. /r/gifs. Similar two-regime pattern.}
\label{fig:gifs_overview}
\end{figure}

\begin{figure}[h]
\centering
\includegraphics[width=0.6\textwidth]{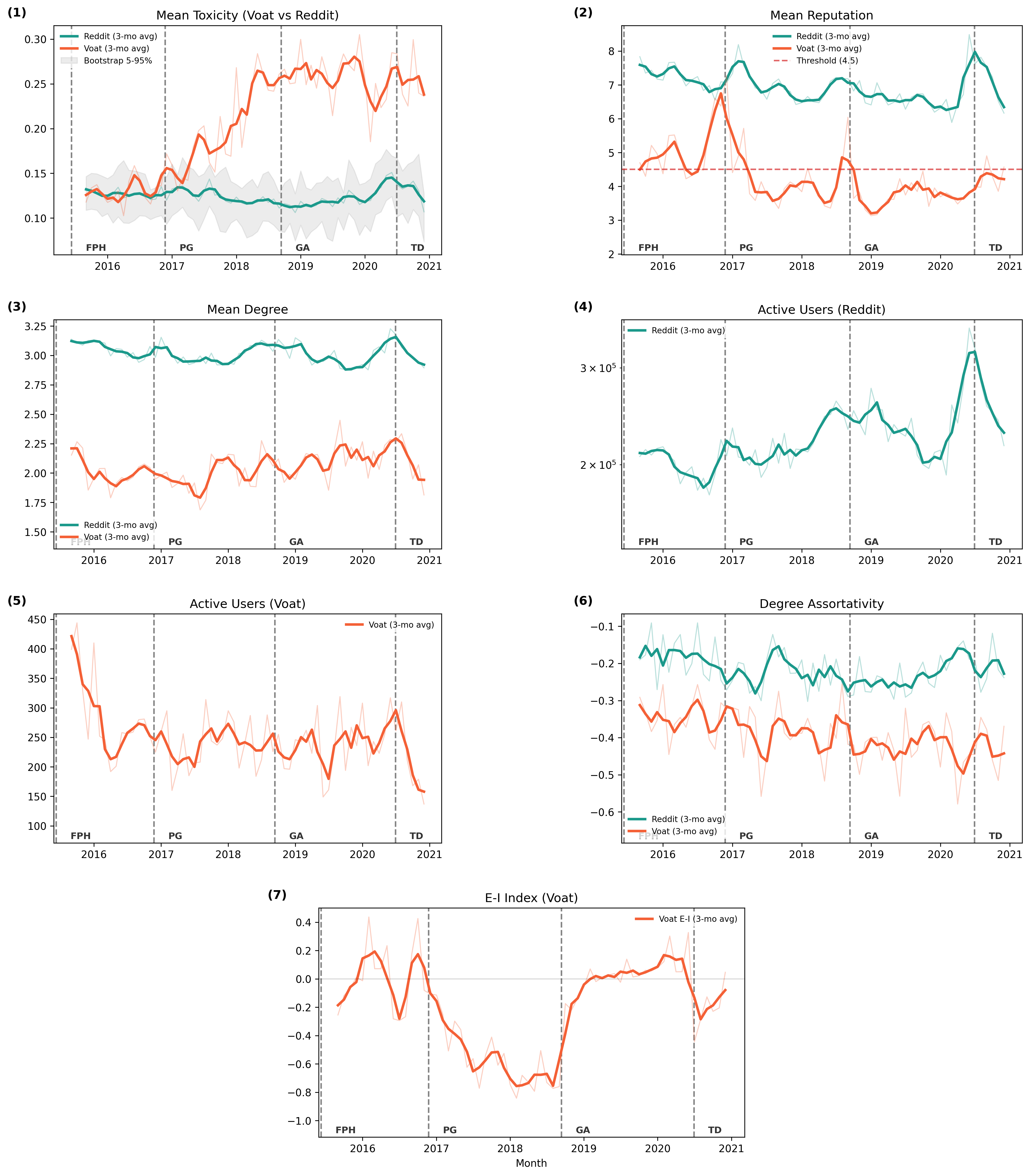}
\caption{Dynamics of /v/pics vs. /r/pics.}
\label{fig:pics_overview}
\end{figure}

\begin{figure}[h]
\centering
\includegraphics[width=0.6\textwidth]{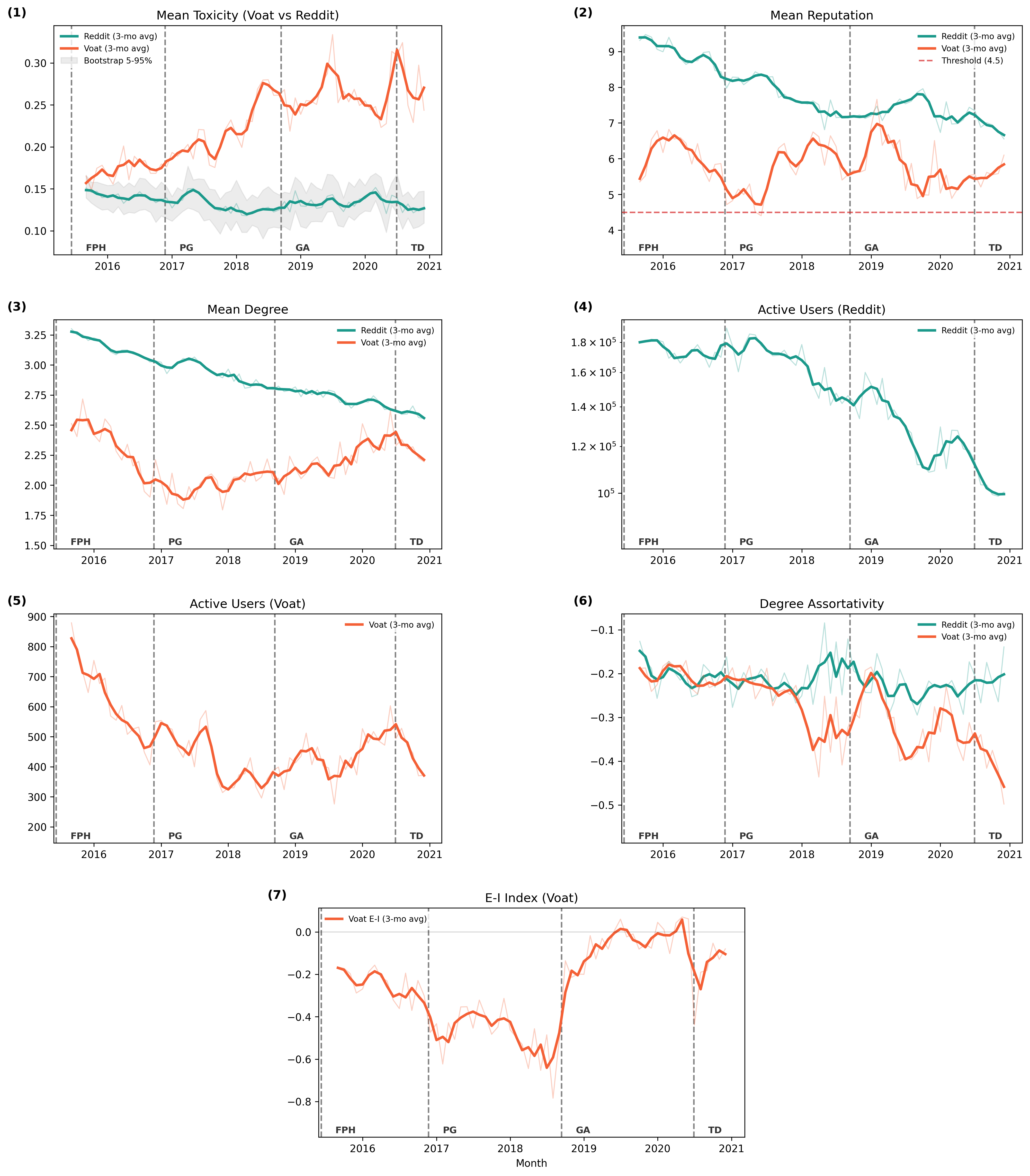}
\caption{Dynamics of /v/videos vs. /r/videos.}
\label{fig:videos_overview}
\end{figure}

\begin{figure}[h]
\centering
\includegraphics[width=0.6\textwidth]{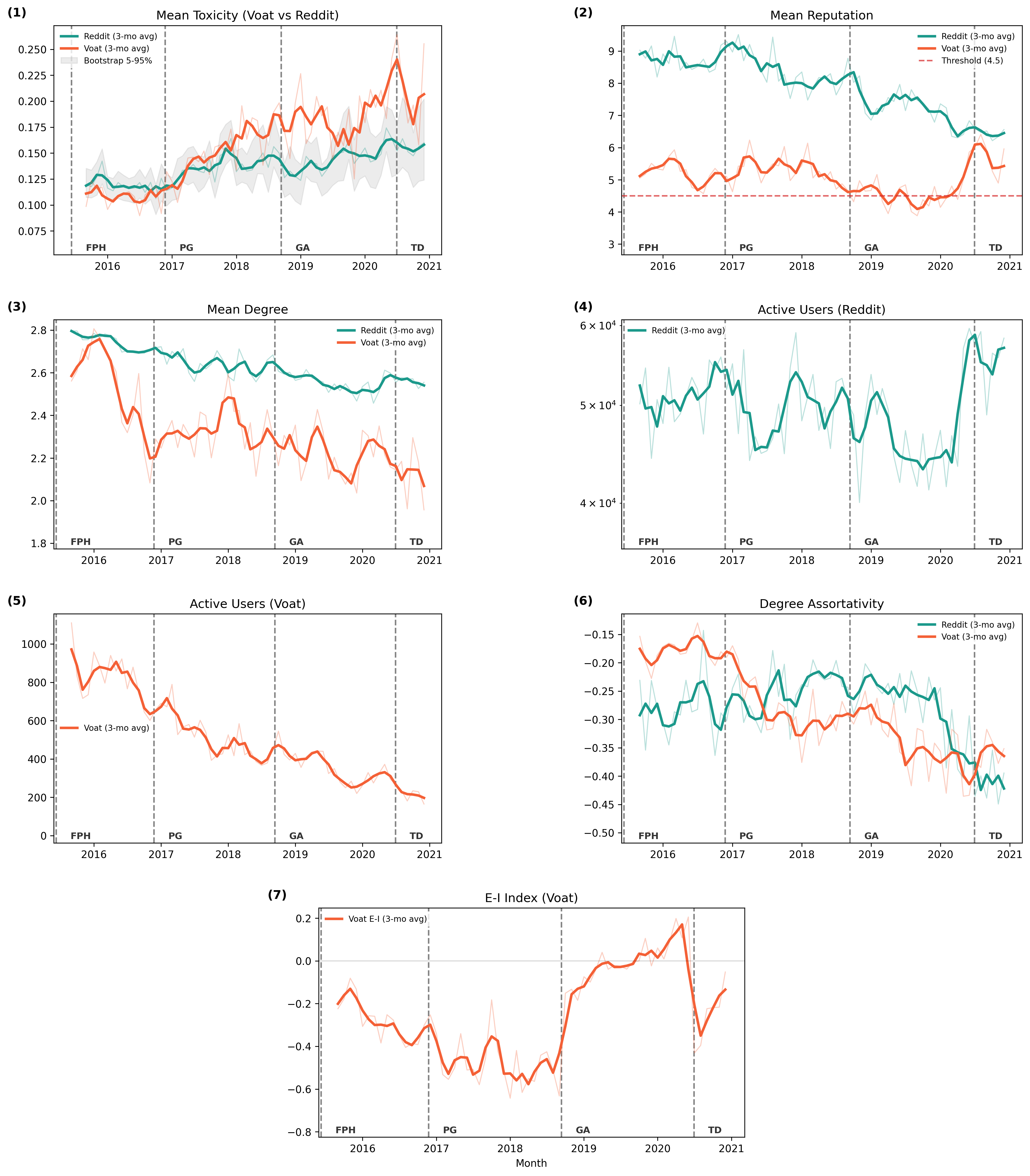}
\caption{Dynamics of /v/technology vs. /r/technology. Attenuated regime patterns, possibly due to topic-focused discussion norms.}
\label{fig:technology_overview}
\end{figure}

\section{Breakpoint Analysis Details}

Table~\ref{tab:breakpoints_full} presents the complete results of the single-break segmented regression analysis. For each community--metric pair, we report the estimated breakpoint month, 90\% bootstrap confidence interval, SSE ratio (lower = stronger break), and bootstrap stability (proportion of 200 iterations within $\pm$3 months of point estimate).

Existing-user Degree Gini is defined here as the Gini coefficient of the degree distribution restricted to Existing users only, used to track changes in hierarchy within the pre-existing core.

\begin{table}[h]
\centering
\caption{Full breakpoint analysis results. CI = 5th--95th percentile bootstrap interval. Stab = bootstrap stability ($\pm$3 months).}
\label{tab:breakpoints_full}
\small
\begin{tabular}{llcccc}
\toprule
\textbf{Community} & \textbf{Metric} & \textbf{Break} & \textbf{CI} & \textbf{SSE Ratio} & \textbf{Stab} \\
\midrule
\multicolumn{6}{l}{\textit{E-I Index (5/7 near GA ban)}} \\
gaming & ei\_index & 2018-10 & [2018-10, 2018-10] & 0.33 & 1.00 \\
pics & ei\_index & 2018-10 & [2018-09, 2018-10] & 0.38 & 1.00 \\
technology & ei\_index & 2018-10 & [2018-10, 2018-10] & 0.36 & 1.00 \\
videos & ei\_index & 2018-10 & [2018-10, 2018-10] & 0.29 & 1.00 \\
global & ei\_index & 2018-10 & [2018-10, 2018-10] & 0.28 & 1.00 \\
funny & ei\_index & 2019-01 & [2019-01, 2019-01] & 0.50 & 1.00 \\
gifs & ei\_index & 2017-02 & [2017-01, 2017-02] & 0.52 & 1.00 \\
\midrule
\multicolumn{6}{l}{\textit{Existing-User Gini (0/7 near GA; all at FPH)}} \\
global & existing\_gini & 2015-07 & [2015-07, 2015-07] & 0.08 & 1.00 \\
funny & existing\_gini & 2015-07 & [2015-07, 2015-07] & 0.13 & 1.00 \\
videos & existing\_gini & 2015-07 & [2015-07, 2015-07] & 0.16 & 1.00 \\
pics & existing\_gini & 2015-07 & [2015-07, 2015-07] & 0.18 & 1.00 \\
technology & existing\_gini & 2015-07 & [2015-07, 2015-07] & 0.23 & 0.99 \\
gaming & existing\_gini & 2015-07 & [2015-06, 2016-01] & 0.33 & 0.83 \\
gifs & existing\_gini & 2015-07 & [2015-07, 2015-09] & 0.43 & 0.99 \\
\midrule
\multicolumn{6}{l}{\textit{Degree Assortativity (0/7 near GA; early 2015)}} \\
technology & degree\_assort. & 2015-07 & [2015-06, 2016-01] & 0.27 & 0.81 \\
videos & degree\_assort. & 2015-05 & [2015-05, 2015-06] & 0.37 & 0.98 \\
global & degree\_assort. & 2015-01 & [2015-01, 2015-02] & 0.41 & 0.99 \\
funny & degree\_assort. & 2015-01 & [2015-01, 2015-01] & 0.46 & 1.00 \\
gifs & degree\_assort. & 2015-09 & [2015-06, 2016-02] & 0.68 & 0.94 \\
gaming & degree\_assort. & 2015-03 & [2015-03, 2019-01] & 0.81 & 0.65 \\
pics & degree\_assort. & 2015-06 & [2015-06, 2018-12] & 0.82 & 0.62 \\
\midrule
\multicolumn{6}{l}{\textit{Mean Toxicity (0/7 near GA; dispersed)}} \\
global & toxicity\_mean & 2018-05 & [2017-12, 2019-05] & 0.50 & 0.61 \\
pics & toxicity\_mean & 2018-04 & [2018-01, 2019-06] & 0.54 & 0.77 \\
videos & toxicity\_mean & 2015-02 & [2015-01, 2015-03] & 0.59 & 0.97 \\
funny & toxicity\_mean & 2015-02 & [2015-02, 2015-02] & 0.60 & 1.00 \\
gaming & toxicity\_mean & 2018-05 & [2018-03, 2018-11] & 0.66 & 0.93 \\
technology & toxicity\_mean & 2019-02 & [2017-05, 2019-04] & 0.79 & 0.77 \\
gifs & toxicity\_mean & 2020-03 & [2017-07, 2020-05] & 0.83 & 0.39 \\
\midrule
\multicolumn{6}{l}{\textit{Mean Reputation (1/7 near GA; dispersed)}} \\
technology & reputation\_mean & 2018-07 & [2018-04, 2018-10] & 0.78 & 0.92 \\
gaming & reputation\_mean & 2016-10 & [2016-10, 2016-10] & 0.40 & 1.00 \\
videos & reputation\_mean & 2016-06 & [2015-10, 2016-07] & 0.64 & 0.91 \\
global & reputation\_mean & 2016-05 & [2015-07, 2016-08] & 0.47 & 0.84 \\
funny & reputation\_mean & 2015-08 & [2015-07, 2015-08] & 0.56 & 1.00 \\
gifs & reputation\_mean & 2014-11 & [2014-11, 2014-12] & 0.73 & 0.97 \\
pics & reputation\_mean & 2015-01 & [2014-12, 2016-02] & 0.83 & 0.89 \\
\bottomrule
\end{tabular}
\end{table}

\section{Robustness Checks}

\subsection{Hub-Threshold Sensitivity}

The main analysis defines hubs as users in the top 10\% of degree distribution. To assess sensitivity to this threshold, we recompute newcomer hub rates at 5\%, 10\%, and 20\% cutoffs. We report `rate vs.\ existing': the proportion of newcomers whose degree exceeds the existing-user hub threshold for that month, which directly tests whether newcomers break into the established elite.

Table~\ref{tab:hub_sensitivity} shows that newcomer hub rates remain low across all thresholds. Even at the most inclusive 20\% cutoff, the global median rate is only 7.1\%, and no community-month observation exceeds 10\% at the global level. The ``no hub capture'' finding is robust to threshold choice.

\begin{table}[h]
\centering
\caption{Hub-threshold sensitivity: newcomer hub rate vs.\ existing-user threshold (global aggregates).}
\label{tab:hub_sensitivity}
\begin{tabular}{lccc}
\toprule
\textbf{Hub Threshold} & \textbf{Median Rate} & \textbf{Max Rate} \\
\midrule
Top 5\% & 0.014 & 0.029 \\
Top 10\% & 0.033 & 0.059 \\
Top 20\% & 0.071 & 0.100 \\
\bottomrule
\end{tabular}
\end{table}

\subsection{Rolling-Tenure E-I Sensitivity}

The main analysis defines newcomers based on first activity after a ban event. An alternative definition uses rolling tenure: users active for fewer than 6 months in any given month are classified as newcomers, regardless of when they first arrived. This captures recency of engagement rather than historical cohort membership.

Table~\ref{tab:ei_rolling} compares pre- and post-GA E-I means under the rolling-tenure definition. The regime pattern persists: all communities show E-I moving toward zero (less segregation) after the GA ban. The global mean shifts from $-0.21$ pre-GA to $-0.08$ post-GA, and /v/gifs achieves positive E-I (+0.02), indicating net cross-cohort interaction. The two-regime finding is robust to alternative newcomer definitions.

\begin{table}[h]
\centering
\caption{Rolling-tenure E-I sensitivity (6-month tenure window). Values are mean E-I indices before and after the GA ban under the rolling-tenure newcomer definition.}
\label{tab:ei_rolling}
\begin{tabular}{lcc}
\toprule
\textbf{Community} & \textbf{Pre-GA Mean} & \textbf{Post-GA Mean} \\
\midrule
funny & $-0.23$ & $-0.17$ \\
gaming & $-0.34$ & $-0.14$ \\
gifs & $-0.23$ & $+0.02$ \\
pics & $-0.10$ & $-0.03$ \\
technology & $-0.20$ & $-0.09$ \\
videos & $-0.17$ & $-0.07$ \\
\midrule
\textbf{Global} & $-0.21$ & $-0.08$ \\
\bottomrule
\end{tabular}
\end{table}

\section{Sentiment Analysis}
\label{sec:sentiment}

We measure sentiment using Valence Aware Dictionary and sEntiment Reasoner (VADER) \citep{hutto2014vader}, a rule-based sentiment analysis tool specifically designed for social media text. VADER uses a lexicon of words with associated sentiment scores and heuristics for punctuation, capitalization, negation, and emojis to determine a compound sentiment score in the range $[-1, 1]$, where $-1$ indicates extremely negative sentiment, $0$ indicates neutral, and $+1$ indicates extremely positive sentiment.

We compute a sentiment score for each comment and post and report mean sentiment per community per month. Sentiment trends are qualitatively similar across communities: both Voat and Reddit communities exhibit mean sentiment values near zero (neutral) with modest fluctuations over time. Unlike toxicity, which shows clear divergence between platforms following ban events, sentiment does not reveal distinctive migration-related patterns.

This finding is consistent with the nature of the metrics: toxicity captures harmful and abusive language specifically, while sentiment measures general positive/negative valence. A community can become more toxic (more hate speech, slurs, and aggressive language) while maintaining neutral overall sentiment if the increased toxicity is balanced by continued neutral or positive non-toxic content. The community overview figures in Appendix A previously included sentiment panels, which have been replaced with summary statistics panels in the current version.

% ============================================================================
% REFERENCES
% ============================================================================

\end{document}